\documentclass[aps,amsfonts,amssymb,amsmath, draft]{revtex4}
\begin{document} 
\title{Kalb-Ramond field interactions in a braneworld
scenario} \author{Ayan Chatterjee\footnote{email: ayan.chatterjee@saha.ac.in}
and Parthasarathi Majumdar\footnote{email:
parthasarathi.majumdar@saha.ac.in}} 
\affiliation{Theory Group, Saha Institute of
Nuclear Physics, Kolkata 700064,  India}
 
\begin{abstract} 
Electromagnetic and (linearized) gravitational interactions of the
Kalb-Ramond (KR) field, derived from an underlying ten dimensional
heterotic string in the zero slope limit, are studied in a five
dimensional background Randall-Sundrum I spacetime with standard model
fields confined to the visible brane having negative tension. The warp
factor responsible for generating the gauge hierarchy in the Higgs
sector is seen to appear inverted in the KR field couplings, when
reduced to four dimensions. This leads to dramatically enhanced
rotation, {\it far beyond observational bounds}, of the polarization
plane of electromagnetic and gravitational waves, when scattered by a
homogeneous KR background. Possible reasons for the conflict between
theory and observation are discussed.
\end{abstract}
\maketitle
\section{Introduction}

The massless antisymmetric tensor field, also known as the Kalb-Ramond
(KR) field, is generic to any closed string spectrum. It is also {\it
not} a degree freedom of the standard low energy theory of fundamental
particle interactions consisting of QCD, the electroweak theory and
general relativity. Any observational effect involving the KR field,
obtained using standard fields as probes, is then a window into the
otherwise inaccessible world of very high energy physics supposedly
predicted by string/M-theories. On the contrary, non-observation of a
large predicted effect is equally likely to illuminate the grey area
where stringy considerations meet the real world.

The weakly coupled heterotic string \cite{gsw} is known to be
consistent and to have $N=1$ supersymmetry, provided the KR 3-form
field strength is augmented by addition of ($E_8 \otimes E_8$)
Yang-Mills Chern-Simons 3-form and local Lorentz Chern-Simons
3-form. This augmentation induces electromagnetic and gravitational
interactions of the KR field which lead to potentially interesting
physical effects showing up in the Maxwell and Einstein equations,
when the theory is compactified to four dimensions. The
electromagnetic effect mainly comprise a rotation of the polarization
plane of electromagnetic waves from large redshift sources, upon
scattering from a homogeneous KR background \cite{ms}-\cite{kmss}.
This rotation is independent of the wavelength of the electromagnetic 
wave and cannot be explained by Faraday effect where the plane of
polarization of the electromagnetic wave rotates depending
quadratically on the wavelength while passing through
galactic/ intergalactic magnetized plasma. 
The magnitude of the effect is sensitive to the dimensional
compactification of the underlying theory. For toroidal
compactification (as well as for the Calabi-Yau compactification) of
the theory (in the zero slope limit), the predicted rotation is
proportional to the appropriate KR field strength component (scaled by
the inverse scale factor in a Friedmann universe), so that bounds on
the observed rotation translate into a stringent upper bound on the
size of the KR field strength component. However, compactifications of
type IIB string theory with p-form fluxes lead to warped spacetimes
\cite{kklt} which lead to rather extraordinary couplings of the KR
field, as we discuss below.

In contrast to the Maxwell field, the gravitational couplings of the KR
field have not been studied substantively in the literature, except in
relation to parity violating KR field couplings \cite{kam}. This paper
addresses this lacuna in the literature. The main finding in this
regard is that gravitational waves exhibit rotation of their plane of
polarization through an angle that is once again proportional to (a
power of) of the KR field strength component. If one uses the bounds
on the KR field strength obtained  from non-observation of the cosmic
optical activity, the effect for  similar activity for gravitational
waves remains small.

The situation changes dramatically in a Randall-Sundrum background of
type I \cite{rs}, i.e., with two three-branes embedded in an
exponentially warped five dimensional anti-deSitter (AdS) spacetime.
In this scenario, the standard model fields are confined to the 
3-brane but gravity can propagate in bulk. Since gravity mode
is a part of the closed 
string spectra, it is not unreasonable to include the closed
string modes like the KR field and dilaton among the $5- d$ fields.
For the case in hand, the dilaton is not significant as it 
couples to the Maxwell Lagrangian and the kinetic term of the KR
field and cannot affect the optical/gravitational activity induced
by the KR- field in any major way. Henceforth, the dilaton will
be freezed to its vacuum expectation value. 
If we assume that (a) the heterotic string admits of compactifications
to such a spacetime (which is then compactified to four dimensions
a\'la ref. \cite{rs}) and (b) the augmented electromagnetic and
gravitational couplings of the KR field survive such compactification,
then the couplings undergo an {\it anti-warping}, i.e., they are
enhanced exponentially, leading to extraordinarily large rotation of
the plane of polarization. Such large-angle rotations of course
contradict observational bounds on these quantities, and hence point
to an area of tension between the set of assumptions made and the real
world. While the exponential enhancement in case of
cosmic optical activity was first derived \cite{mais}
through an analysis involving the five dimensional action, in this
paper we provide a more direct approach to this where the interplay
between the equations of motion and the Bianchi identities becomes
manifest.

The plan of the paper is as follows: In section 2, we briefly review
our original derivation of the Cosmic Optical Activity and extend it
to the case of gravitational waves from distant sources. In section 3,
the anti-warped coupling of the KR field to electromagnetism is
obtained by manipulating the field equations and Bianchi identities in
the warped spacetime and effecting the compactification to four
dimensions. This is repeated in section 4 for gravitational waves.  We
conclude in section 5 with an attempt to analyze possible reasons for
the conundrum.


\section{Physical effects of KR interactions}

The KR field is characterized by a 2-form potential $B$ which has a
3-form field strength $H \equiv dB$; the field strength is invariant
under the KR gauge transformation $\delta_\Lambda B~=~d \Lambda$,
where $\Lambda$ is a one-form gauge parameter. The free KR action is
given (in $D$ dimensional spacetime) as
\begin{eqnarray}
S_{H}~=~\int_{{\cal M}_D} ~H~\wedge ~^*H~, \label{kract}
\end{eqnarray}
where, $^*H$ is the Hodge-dual of the field strength $H$. Varying this
action w.r.t. $B$ yields the KR field equation
\begin{eqnarray}
d^*H~=~ 0 \label{kreom}
\end{eqnarray}
which has the immediate local solution
\begin{eqnarray}
^*H~=~d V, \label{krsol}
\end{eqnarray}
where, $V$ is a $D-4$ form. Substituting this in the KR Bianchi
identity
\begin{eqnarray}
dH ~=~0 \label{krbi}
\end{eqnarray}
one obtains for the field $V$
\begin{eqnarray}
d ^*dV ~=~0~. \label{boxv}
\end{eqnarray}

Ten dimensional heterotic string theory (where $B$ occurs in the
massless spectrum of the free string) reduces in the zero slope
(infinite tension) limit to ten dimensional $N=1$ supergravity coupled
to $N=1~ E_8 \otimes E_8$ super-Yang-Mills theory. The requirement of
ten dimensional supersymmetry requires that the KR field strength $H$
be augmented as \cite{gsw}
\begin{eqnarray}
H~=~dB~-~M_D^{1-D/2}~\Omega_{YM} ~,\label{krym}
\end{eqnarray}
where
\begin{eqnarray}
\Omega_{YM}~\equiv~tr(A \wedge dA~+~\frac23 g A\wedge A  \wedge A)
~\label{ymcs}
\end{eqnarray}
is the Yang-Mills Chern-Simons 3-form with $A$ the gauge  connection
1-form and $ M_{D}$ is the Planck mass in $D$- dimensional spacetime.
The theory in this form still suffers
from the  presence  of mixed anomalies at the quantum level;
consistency is restored upon a further augmentation \cite{gsw}
\begin{eqnarray}
H~=~dB~-~M_D^{1-D/2}~(\Omega_{YM}~-~\Omega_{L}) ~,\label{krfu}
\end{eqnarray}
where $\Omega_{L}$ is the gravitational Chern-Simons 3-form  obtained
by replacing the Yang-Mills gauge connection $A$ by the  spin
connection 1-form $\omega$ in (\ref{ymcs}), and the trace is  taken
over the local Lorentz indices.

The augmentation in eq. (\ref{krfu}) has the following consequences:
\begin{itemize}
\item In order that $H$ remain gauge invariant under both Yang-Mills
gauge transformations and under local Lorentz transformations, $B$
must  now transform non-trivially under both gauge transformations.
This is a surprise as the field $B$ is neutral and also has no
magnetic moment. We consider the gauge transformation of the Yang-Mills field
$A$, given by
\begin{eqnarray}
\delta_{YM} A ~=~ d\Sigma ~+~ [A,\Sigma] \label{ymgt},
\end{eqnarray}

where, $\Sigma$ is a matrix of infinitesimal parameters. 
The Chern-Simons term varies as 
\begin{eqnarray}
\delta_{YM} \Omega_{YM} ~=~ tr(d \Sigma~ \wedge  dA)\label{ymcsvar}
\end{eqnarray}
Thus, to achieve gauge invariance for the $H$ field, the
transformation law for $B$ should include the $2$-form in (\ref{ymcsvar})
so that under Yang-Mills gauge transformation
\begin{eqnarray}
\delta_{YM} B = M_D^{1-D/2}tr(d \Sigma~\wedge dA) \label{ymb}
\end{eqnarray}

Also, the gravitational field in the vielbein formalism can
be treated very similarly to the Yang-Mills field. Specifically
the Yang-Mills potential $A$ is analogous to the spin connection
1-form $\omega_{AB}$, where $A, B$ are Lorentz indices. Under an
infinitesimal Lorentz transformation with parameters given by
an $SO(D- 1,1)$ matrix $\Theta$, the transformation of $\omega$
is
\begin{eqnarray}
\delta_{L} \omega  ~=~ d\Theta ~+~ [\omega,\Theta] \label{lgt},
\end{eqnarray}

The Lorentz Chern-Simons term varies as 
\begin{eqnarray}
\delta_{L} \Omega_{L} ~=~ tr(d \Theta~ \wedge d\omega)\label{lcsvar}
\end{eqnarray}
Similar to the argument above, transformation law for $B$
should include the $2$-form in (\ref{lcsvar})
so that under Lorentz transformation
\begin{eqnarray}
\delta_{L} B = - M_D^{1-D/2}tr(d \Theta~\wedge d\omega) \label{lmb}
\end{eqnarray}

\item Retaining the form of the KR action (\ref{kract}), it follows
that the KR field equation does not change. Therefore, $^*H$  still
has the local solution (\ref{krsol}). However, the KR Bianchi identity
certainly changes, leading to
\begin{eqnarray}
d ^*dV~=~M_D^{1-D/2}~tr(F \wedge F~-~R \wedge R) ~,\label{krbii}
\end{eqnarray}
where $F(R)$ is the Yang-Mills (spacetime) curvature 2-form.
\item The Yang-Mills and Einstein equations change non-trivially. We
shall consider these below in special situations viz., the Maxwell
part of the gauge interaction and linearized gravity.
\end{itemize}


\subsection{Cosmic optical activity}

In this subsection we confine ourselves to the electromagnetic
interactions of the KR field after toroidally compactifying the
heterotic string in the zero slope limit to four dimensional Minkowski
spacetime. The relevant four dimensional field equations  are
\begin{eqnarray}
\partial_{\mu} H^{\mu \nu \rho}~& =&~0~ \nonumber \\ \partial_{\mu}
F^{\mu \nu}~&=&~ M_P^{-1}~H^{\nu \rho \eta}~F_{\rho  \eta}
~. \label{eom}
\end{eqnarray}
The corresponding Bianchi identities are
\begin{eqnarray}
\Box \Phi_H ~&=&~ M_P^{-1}~F^{\mu \nu}~^*F_{\mu \nu} \nonumber \\
\partial_{\mu} ^*F^{\mu \nu}~&=&~0 ~, \label{bia}
\end{eqnarray}
where, in four dimensions, the (pseudo)scalar $V=\Phi_H$ and $M_{P}$ 
is the Planck mass in $4d$-spacetime.

Rather than solving these equations simultaneously, we introduce
another simplifying assumption: the `axion' field $\phi_H$ is {\it
homogeneous} and provides a background with which the Maxwell field
interacts. We restrict our attention to lowest order in the inverse
Planck mass $M_P$, so that terms on the RHS of the axion field
equation (\ref{bia}) are ignored to a first approximation.
Consequently, ${\dot \Phi}_H \equiv d\Phi_H/dt = f_0$ where  $f_0$ is
a constant of  dimensionality of $(mass)^2$. Under these conditions,
the Maxwell  equations can be combined to yield the inhomogeneous wave
equation  for the magnetic field ${\bf B}$
\begin{eqnarray}
\Box {\bf B}~=~-~{2~f_0\over M_{P}}~{\bf \nabla \times B} ~. \label{wave}
\end{eqnarray}
With the ans\"atz for a plane wave travelling in the  $z$-direction,
${\bf B}({\bf x},t)= {\bf B}_0(t)~\exp  ikz$, we obtain, for the
left and the right 
circular polarization states $B_{0 \pm}  \equiv B_{0 x} \pm i B_{0 y}$,
\begin{eqnarray}
{d^2 B_{0 \pm} \over dt^2}~+~(k^2 ~\mp~ {2 f_0 k \over M_P})~B_{0
\pm}=~0  ~.\label{poleq}
\end{eqnarray}
Similarly, we can obtain the wave equation for the left and 
right circularly polarization states for the electric field
which has exactly the same form as that of magnetic field.
We concentrate on the equation for magnetic field as the conclusions
will be same for that of electric field. 
Thus, the right and left circular polarization states have different
angular frequencies (dispersion)
\begin{eqnarray}
\omega_{\pm}^2 = k^2 ~\mp~ {2k f_0 \over M_P} ~\label{disp}
\end{eqnarray}
so that over a time interval $\Delta t$, the plane of polarization
undergoes a rotation (for large $k$)
\begin{eqnarray}
\Delta \Psi_{op} \equiv |\omega_+ - \omega_-|~\Delta t ~\simeq ~2  {f_0
\over M_P}~\Delta t ~.\label{rota}
\end{eqnarray} 

In a radiation or matter dominated Friedmann universe, the formula for
the angle of rotation changes slightly \cite{kmss}
\begin{eqnarray}
\Delta \Psi_{op} \equiv |\omega_+ - \omega_-|~\Delta t ~\simeq ~2  {f_0
\over a^2(t) M_P}~\Delta t ~, \label{rotf}
\end{eqnarray} 
where, $a(t)$ is the scale factor and $\Delta t$ is now to be taken as
the look-back time. This means that $\Delta \Psi = \Delta \Psi(z)$, 
where $z$ is the red-shift, and increases with red-shift. This rotation
also differs from the better-understood Faraday rotation in that it is
{\it achromatic} in the limit of high frequencies, \footnote{We have
made an assumption here which is shared with Faraday rotation:  the
existence of a coherent electromagnetic wave  over a time $\Delta
t$. Cosmologically, any vector  perturbation tends to thermalize with
time scales typically smaller  than $\Delta t$. We thank
T. Padmanabhan for pointing this out to  us.}. Observationally, even
for large redshift sources, the angle of  rotation is less than a
degree, which imposes the restriction on the  dimensionless quantity
$f_0/M_P^2 < 10^{-20}$, \footnote{In regard to astrophysical
observations of optical activity, it appears that there is definite
evidence that the rotation of the plane of polarization travelling
over cosmologically large distance is not entirely attributable
to Faraday rotation due to magnetic fields present in the galactic
plasma \cite{JR}. It is therefore not unlikely that the axion field will
endow observable effect in CMB.}, \footnote{A variant  of the above KR-Maxwell
interaction with parity violating  effects, not obviously grounded in
string/M-theory, has been shown  to lead to parity violating
correlations  between temperature and polarization anisotropy in the
Cosmic Microwave Background \cite{pm}.}.


\subsection{Cosmic gravitational activity} 

To find the gravitational analogue of the optical activity discussed
above, we first note that the augmentation of $H$ in (\ref{krfu})
implies that the $tr R \wedge R$ term contributes an additional term
to the Einstein equation over and above the energy-momentum  tensor of
the KR field. Formally,
\begin{eqnarray}
{\cal G}_{\mu \nu} ~=~{8\pi \over M_P^2}T_{\mu \nu} ~+~{16\pi  \over
M_P^3}~{1 \over \sqrt{-g}}~{\delta  \over \delta g^{\mu \nu}}~\int
d^4x'~\sqrt{-g}(x')~\Phi_H(x')~  R_{\rho \lambda \sigma
\eta}(x')~^*R^{\rho \lambda \sigma \eta}(x')~,
\label{ee}
\end{eqnarray}
where,
\begin{eqnarray}
T_{\mu \nu}~=~H_{(\mu| \tau \rho}~H_{\nu)}{}^{\tau \rho}~-~\frac16
g_{\mu \nu} H^2 ~. \label{tmn}
\end{eqnarray}

Since our focus is on gravitational waves, it is adequate to  consider
the Einstein equation in a linearized approximation. To this effect,
we decompose the metric $g_{\mu \nu} = \eta_{\mu \nu} +  h_{\mu \nu}$
with the fluctuation $h_{\mu \nu}$ being considered  small so that one
need only retain terms of $O(h)$ in the Einstein  equation. We further
impose on the fluctuations $h_{\mu \nu}$ the Lorenz gauge
$h_{\mu \nu},{}^{\nu} = \frac{1} {2}  h,{}_{\mu} $ .

In this gauge, the
linearized Einstein equation becomes
\begin{eqnarray}
-\Box~ h_{\mu \nu}~=~{16\pi \over M_P^2}~T_{\mu \nu}~-~{128\pi \over
  M_P^3}~\epsilon_{(\mu|}^{~~~\sigma \alpha \beta}~\left[\Phi_{H,\lambda
  \sigma}~\left(h_{\beta |\nu), \alpha}^{~~~~~~\lambda}~+
  ~h^{\lambda}_{~\beta,\alpha |\nu)} \right) ~- ~\Phi_{H,\alpha}~\Box
  h_{\beta |\nu),\sigma} \right] ~ \label{linee}
\end{eqnarray}

In analogy with the last subsection, we regard the axion field
$\Phi_H$ as a homogeneous background satisfying eq. (\ref{bia}) and
consider its effect on a plane gravitational wave. We restrict to  the
lowest inverse power of the Planck mass for which a nontrivial  effect
is obtained, and as such, ignore terms on the RHS of the  axion field
equation. 
\begin{eqnarray}
\Box \Phi_H ~&=&~ M_P^{-1}~R^{\mu \nu \lambda \sigma}~^*R_{\mu \nu
\lambda \sigma }
\end{eqnarray}


Now, the field equations are (before gauge choice) invariant
under general coordinate
transformation. As we have chosen the Lorenz gauge, 
this invariance is broken and not all components
of  $h_{\mu \nu}$ are independent.
In fact, the only  physical degrees
of freedom of the spin 2 field are contained in  $h_{ij}$, for which
we choose a plane wave ans\"atz travelling in the z- direction,
\begin{eqnarray}
h_{ij}~=~\varepsilon_{ij}(t)~\exp ~-ikz ~. \label{gw}
\end{eqnarray}

The Latin indices above correspond to spatial directions. The other
components of $h_{\mu\nu}$ can be gauged away, so that  their field
equation need not be considered.
The only non-vanishing polarization components can be chosen to be
$\varepsilon_{11} = -~\varepsilon_{22}~,~\varepsilon_{12}=
\varepsilon_{21}$; from these the circular polarization components can
be constructed as in the Maxwell case: $\varepsilon_{\pm} \equiv
\varepsilon_{11} \pm i \varepsilon_{12}$. Further, 
we write the energy momentum tensor in eq. (\ref{tmn}) in terms of
$\Phi_{H}$ using eq. (\ref{krsol}). Then, under the approximation
of homogeneous axion field, these polarization
components satisfy the  inhomogeneous differential equation
\begin{eqnarray}
\left[{d^2 \over dt^2}~+~k^2 ~+~{\cal F}_{\pm} \right]
\varepsilon_{\pm}~=~-~{\cal F}_{\pm}~, \label{dife}
\end{eqnarray}
where,
\begin{eqnarray}
{\cal F}_{\pm}~\equiv~{8\pi f_0^2 \over M_P^2~(1 \pm  {128\pi k f_0
/M_P^3}) } ~ . \label{eff}
\end{eqnarray}
The difference between (\ref{dife}) and the analogous equation
(\ref{poleq}) is that the former has a forcing term absent in the
latter; this forcing term is dependent on the wave number $k$ and
controlled by the constant $f_0$ which characterizes the strength of
the KR field coupling.

Even though we are interested in large $k$, we would still remain
within the Planckian regime $k < M_P$ so that the quantity $16\pi k
f_0/M_P^3 << 1$ and can serve as an expansion parameter, leading to
\begin{eqnarray}
\left[{d^2 \over dt^2} ~+~ k^2 ~+~ 8\pi f_0^2/M_P^2  ~\mp~1024\pi^2 k
f_0^3 /M_P^5 \right] \varepsilon_{\pm}~\simeq~-~8\pi f_0^2~(1 \mp 16
\pi k f_0/M_P^3)/M_P^2~ . \label{diffe}
\end{eqnarray}
We can now read off the dispersion relation
\begin{eqnarray}
\omega_{\pm}^2 ~ = ~ k^2 ~+~ 4\pi f_0^2/M_P^2  ~\mp~1024\pi^2 k f_0^3
/M_P^5 \label{disp}
\end{eqnarray}
whence the group velocity
\begin{eqnarray}
v_{g \pm}~ \equiv ~ {d\omega_{\pm} \over dk} ~=~ 1 ~+~ O(k^{-2})~,
\label{gvel}
\end{eqnarray}
and the phase velocity
\begin{eqnarray}
v_{p \pm}~ \equiv ~ {\omega_{\pm} \over k} ~=~ 1 ~\mp~512 \pi^{2}
f_0^3/M_P^5k ~.\label{pvel}
\end{eqnarray}
Thus for large $k$ the violation of Lorentz invariance can be
ignored. As in the electromagnetic case, the rotation of the
polarization plane for gravitational waves is given by
\begin{eqnarray}
\Delta \Psi_{grav}~\simeq ~1024 \pi^{2} {f_0^3 \over M_P^5}~\Delta t ~.
\label{grota}
\end{eqnarray}
Admittedly, the effect is immeasurably tiny; with the limits on $f_0$
given in the previous subsection, it is $O(10^{-30})$. However, in
contrast to the optical case, the tensor perturbations characterizing
the gravitational wave do not get randomized, so the effect is in
principle observable.


\section{KR-Maxwell interactions in an RSI Braneworld}

Using eq. (\ref{krbii}), and retaining only the coupling to the
Maxwell field, the KR Bianchi identity in a five dimensional RS1
background \cite{rs}
\begin{eqnarray}
ds^2 = e^{-2 \sigma(y)}~ \eta_{\alpha \beta}~ dx^{\alpha}  dx^{\beta} ~+~
dy^2~, \label{rsmet}
\end{eqnarray}
is given by
\begin{eqnarray}
\Box_{RS} V^M ~=~ M_5^{-3/2}~^*F^{MNP}~F_{NP} ~, \label{rskr}
\end{eqnarray}
where, $\Box_{RS}$ is the covariant d'Alembertian on five dimensional
warped RS1 spacetime, $M,N,P,= 0, 1,\dots, 4$,~ $\sigma(y)=~ 
k|y|$,~ $F_{NP} \equiv \partial_{[N} A_{P]}$ and~ $^*F^{MNP}$
is the tensor derived using the antisymmetric tensor in RS1
spacetime.  Since the Maxwell field is supposed to be confined to the
(visible) brane, $A_4~ =~ 0$, and also $A_{\mu}$ are independent of the
compact coordinate $y$, so that $F_{4\mu}~ =~ 0$. Recalling that the
five-vector $V^M =~ ( V^{\mu},\Phi_H)$, the axion field
$\Phi_H(x,y)$ satisfies
\begin{eqnarray}
\Box_{RS} \Phi_H ~=~M_5^{-3/2}~^*F^{\mu \nu }~F_{\mu
\nu}~\delta(y~-~a) ~. \label{max}
\end{eqnarray}
The corresponding vector equation is 
\begin{eqnarray}
\Box_{RS} V^{\mu} ~= 0 \label{vecmax}
\end{eqnarray}
Using the metric (\ref{rsmet}), eq. (\ref{max}) can be rewritten
\begin{eqnarray}
e^{2\sigma(y)}~\left[ \Box\Phi_H~+ ~~e^{2\sigma(y)}
 \partial_{y} \{e^{-4 \sigma(y)}~ \partial
_{y}\Phi _H \}~\right] =~M_5^{-3/2} ~ g^{\lambda \mu}g^{\sigma \nu}~\tilde
F_{\lambda \sigma }~F_{\mu \nu}~\delta(y~- ~a)  ~, \label{maxx}
\end{eqnarray}
where, $\Box$ is the d'Alembertian on four dimensional Minkowski
space, and $y~\in~ [0,~a]$ is the coordinate on the fifth
dimension. Also,  we have made use of the transformation properties of
complete antisymmetric tensor from the RS spacetime to that of the
Minkowski spacetime. The $\tilde F^{\mu \nu}$  tensor is that as
observed in Minkowski spacetime.

The compactification to four dimensions now proceeds through the
ans\"atz
\begin{eqnarray}
\Phi_H(x,y)~=~\sum_n~\Phi_H^{(n)}(x)~\chi_n(y)~, \label{deco}
\end{eqnarray}
where, the mode functions $\chi_n(y)$ satisfy the eigenvalue
equation (for each value of $n$)
\begin{eqnarray}
{d \over dy}\left[e^{-4\sigma(y)}~{d\chi_n(y) \over
dy} \right]~=~m_n^2~e^{-2\sigma(y)}\chi_n(y)
~. \label{eig}
\end{eqnarray}

So that the equation (\ref{maxx}) reduces to
\begin{eqnarray}
\sum_n e^{-2\sigma(y)}~\left[ \Box\Phi_H^{(n)}(x)~+~~m_n^2\Phi
_H^{(n)}(x) ~\right]\chi_n(y) =~M_5^{-3/2} ~~\tilde F^{\mu \nu
}~F_{\mu \nu}~\delta(y~-~a)  ~, \label{eomphi}
\end{eqnarray} 
Our interest here is in the zero-mode $\chi_0(y)$ corresponding
to $m_0=0$. It is easy to verify that the second order differential
operator in eq. (\ref{eig}) is self-adjoint in the domain $[0,~a]$
provided
\begin{eqnarray}
\chi_0(0)~&=&~ {1\over\alpha}~{d\chi_0(y) \over
dy}|_{y=0} ~ \nonumber \\
\chi_0(a)~&=&~{1\over\beta}~{d\chi_0(y) \over
dy}|_{y = a} ~ \label{sad}
\end{eqnarray}
where $\alpha,\beta$ are real constants having value $4k$. The
eigenfunctions $\chi_n (y)$ satisfy the  orthonormality condition

\begin{eqnarray}
\int_{0}^{a} {e^{-2\sigma(y)}}\chi_{m}(y) \chi_{n}(y)~
dy = \delta_{m n} ~\label{ortho}
\end{eqnarray}

Choosing an ans\"atz for the equation (\ref{eig}),~ $\chi_0(\theta)=C_1
\exp 4\sigma(y) + C_2$, the boundary condition (\ref{sad})
implies $C_2 = 0$. Using (\ref{ortho}), we obtain the normalized
solution~ $\chi_0(y) = \sqrt{6k}  \exp k(4y~
-~3a)$. Substituting this solution in equation (\ref{eomphi}), and
using the orthonormality of the eigenfunctions in (\ref{ortho}), one
obtains the four dimensional equation of motion for the zero mode KR
axion field
\begin{eqnarray}
\Box \Phi_H^{(0)}(x)~=~{\exp \sigma(a) \over M_P}~F^{\mu \nu} \tilde
F_{\mu \nu} ~,
\label{axeq} 
\end{eqnarray}
where, $M_P = (M_5^{3}/k)^{1 \over 2}[1- e^{-2\sigma(a)}] \simeq
(M_5^{3}/k)^{1 \over 2}$ is the Planck mass in $4- d$
spacetime. Notice
that the warp factor in the RS metric
(\ref{rsmet}) now appears inverted, implying an {\it exponentially
large enhancement in the KR-Maxwell coupling strength}~ \footnote{It is
easy to verify that the remaining components of the 5-dimensional dual
KR gauge field $V^M$ decouple from the Maxwell field on the brane and
need no longer be considered.}.

The mode solution for the field $V^{\mu}(x, y)$ can be calculated similarly.
The field $V^{\mu}(x,y)$ when expanded in the Kaluza-Klein modes

\begin{eqnarray}
V_{\mu}(x,y)~=~\sum_n~V_{\mu}^{(n)}(x)~\zeta_n(y)~, \label{vecdeco}
\end{eqnarray}

leads to the mode equation (for each mode)

\begin{equation}
\frac{d}{d y} \left(e^{- 2 \sigma(y)} 
\frac{d}{dy} \, \zeta^{n}(y)\right) = m_n^2 \, \zeta^{n}(y)\,.
\label{Eigen}
\end{equation}

The orthonormality condition for $\zeta(y)$ is 
\begin{equation}
\int_{0}^{a}  \, \zeta_{m}(y)~ \zeta_{n}(y)~dy = \delta_{m n} 
\label{Norm}
\end{equation}
It is easy to verify using (\ref{Eigen}) and (\ref{Norm})
that the zero mode is a constant on the visible brane.

\vskip 0.5 cm
Analysis similar to (\ref{rskr}) leads to the Maxwell equation on
the visible brane: one starts with the equation
\begin{eqnarray}
\partial_{\mu} F^{\mu \nu}~=~M_5^{-3/2}~ \tilde F^{\nu
\lambda}~[\partial_{\lambda} \Phi_{H}~-~\partial_y V_{\lambda}]
~\label{fullmaxkr}.
\end{eqnarray}
Expanding the dual KR vector field in modes and using its dynamics,
the zero mode of $V_{\lambda}$ can be shown to decouple, yielding
\begin{eqnarray}
\partial_{\mu} F^{\mu \nu}~=~{\sqrt{6}\exp \sigma(a) \over M_P}~ \tilde F^{\nu
\sigma}~\partial_{\sigma}\Phi_H^{(0)}(x)~. \label{maxkr}
\end{eqnarray}
Once again, we observe the exponentially enhanced coupling strength
seen in the effective four dimensional field equation for the axion
(\ref{axeq}). This enhancement in the effective interaction strength
between the KR and Maxwell fields was first seen in ref. \cite{mais}
using a compactification analysis of the effective action. Here, we
have preferred to re-derive it using direct compactification of the
field equations {\it and} Bianchi identities in the bulk, underlining
thereby the crucial role played by the latter which are not directly
derived from the action. In this manner, the result of
ref. \cite{mais} is placed on a firmer footing.

The exponentially large coupling translates immediately into an
exponential increase in the size of the angle through which the
polarization plane of an electromagnetic wave, interacting with a
homogeneous sea of KR axions, rotates \cite{mais}.
\begin{eqnarray}
\Delta \Psi_{op} \equiv |\omega_+ - \omega_-|~\Delta t ~\simeq
~{2\sqrt{6}  f_0~e^{ka}
\over M_P}~\Delta t ~.\label{rsrota}
\end{eqnarray}

Now, we had denoted the strength of the homogeneous axion field
in flat spacetime as $f_{0}$. In $RS$ spacetime, this will convert to
$\tilde f_{0}=~ f_{0}e^{ka}$ on the negative tension brane.
So, essentially, the strength
of the axion field gets enhanced by an anti-warping factor $e^{ka}$.
Thus, the rotation of the plane of polarization is
\begin{eqnarray}
\Delta \Psi_{op} \equiv |\omega_+ - \omega_-|~\Delta t ~\simeq
~{2\sqrt{6}  \tilde f_0~
\over M_P}~\Delta t ~.\label{rsrota}
\end{eqnarray}

The cosmic optical
activity thus is far in excess of anything that is, or will be, ever
observed from cosmologically distant galaxies, and is therefore
physically untenable. One may argue, as already mentioned, that the
incompatibility between theory and observation seen so far is actually
itself rather unlikely from a cosmological standpoint, since vector
perturbations of the metric (e.g., electromagnetic waves) would
certainly thermalize like the CMBR on time scales shorter than those
under consideration. Note however that the same enhancement shows up
in a parity-violating modification leading to exponentially large
$B$-type parity-violating  polarization anisotropy of the CMBR itself
\cite{cowh}, which is certainly  within the realm of
observability. As we show below, the same is true for tensor
(i.e., gravitational) perturbations, where thermalization
is not effective in washing out long wavelength effects.
An exponential coupling to KR fields would
lead to an unobservably large effect on the polarization of
gravitational waves, far beyond what one expects to see in
gravitational wave experiments like LIGO and LISA.


\section{KR- gravitational interactions in an RSI Braneworld}
 
The five dimensoinal RS metric \cite{rs}, given by eq. (\ref{rsmet})
is subjected to linear perturbations
\begin{eqnarray}
ds^{2}= dy^{2}+ (e^{- 2\sigma(y)}~\eta_{\mu \nu} + h_{\mu \nu})~
 dx^{\mu}dx^{\nu}
\end{eqnarray}
$i.e. ~ g_{MN}=~\stackrel{\circ}g_{MN}+~ h_{MN}$. these pertubations
further obey the $RS$ gauge conditions \cite{rs},
\begin{eqnarray}
h_{55}= h_{\mu 5}= 0 \nonumber\\ h_{\mu \nu},{}^{\nu}=~0 ,
\nonumber\\ h_{\mu}^{\mu} =~ 0~. \label{rsgau}
\end{eqnarray}
In what follows, we retain terms upto $O(h)$ in the field
equations. 

To obtain a degree of freedom count, observe that the totality of $10$
gauge conditions in (\ref{rsgau}) leave $5$ independent
components of the spin 2 graviton tensor in five dimensional
spacetime. These degrees of freedom are further split
into  $4$ dimensional graviton ($2$ polarizations),
a $4$ dimensional spin 1 photon ($2$
polarizations) and a spin zero scalar, also in four dimensions.
In the general case of $n$ extra dimensions, the number
of degrees of freedom of graviton follows from the 
irreducible tensor representations of the isometry
group as ${1\over 2} (n~+~1)(n~+~2)$. The massive modes of the $ 5$
dimensional gravitons are represented by the massive modes
of all these  fields on
the brane. Since we will be concerned with the tensor perturbations,
only the $h_{ij}$ is of interest. The standard $4$ dimensional graviton
corresponds to the massless mode of $h_{ij}$.

We look for the solution, to $O(h)$,  of the five dimensional Einstein
equation
\begin{eqnarray}
^{(5)}G_{MN} \equiv {}^{(5)}R_{MN} - {1 \over 2} {}^{(5)}g_{MN}
{}^{(5)} R = - \Lambda_{5}{}^{(5)}g_{MN} + {}^{(5)}\kappa T_{MN}
\end{eqnarray}
where, $\Lambda = -6k^{2}$ is the cosmological constant of the $AdS_5$
spacetime, ${}^{(5)}\kappa = 8\pi G_{5}$, $G_{5}=~ 1/ M_{5}^{3}$  
is the five dimensional
Newton's constant and $T_{MN}$ is the energy -momentum tensor which is
the sum of energy-momentum of the KR field and the term obtained from the
variation of the Chern-Simons- Kalb-Ramond field $i.e$
\begin{eqnarray}
&&T_{M N}~=~t_{M N} ~+~{1 \over M_5^{3\over2}}~{1 \over
\sqrt{-g}}~{\delta  \over \delta
g^{MN}}~\int d^5x'~\sqrt{-g}(x')~V_{P}(x')~
R_{QRST}(x')~^*R^{PQRST}(x')~,
\end{eqnarray}
where,
\begin{eqnarray}
&&t_{MN}~=~H_{(M|TR}~H_{N)}^{T R}~-~\frac16 g_{MN} H^2 ~ \label{emtensor}
\end{eqnarray}





The aim here is to obtain the leading order coupling of the induced
four dimensional graviton fluctuations on the visible brane, with the
Kalb-Ramond field,
in order to ascertain the effect of the latter on the polarization of
(to be) observed gravitational waves. In this venture, we are guided
by the earlier work of ref. \cite{GT}, with the input of the
appropriate energy momentum tensor involving the antisymmetric tensor
fields.

Proceeding as in \cite{GT}, the linearized Einstein equation in $RS$
gauge reduces to
\begin{equation}
\left[e^{2\sigma(y)}~\Box^{(4)} + \partial_y^2 -4k^{2}+ 4k
(\delta(y)+\delta(y-a) )\right] h_{\mu\nu}=-2\kappa
\Sigma_{\mu\nu}\delta(y-a)\label{rseomgr}
\end{equation}
where,
\begin{equation}
\Sigma_{\mu\nu}= \left( {T_{\mu\nu}-{1\over 3} \gamma_{\mu\nu} T}
\right)  +2 \kappa^{-1} \xi^5_{,\mu\nu}~.  \label{impem}
\end{equation}
Here, the delta functions in (\ref{rseomgr}) will enforce the
discontinuities. The combination in (\ref{impem}), includes the
``bending'' of the wall $\xi^5$, and will
play the role of the source term in the RS gauge. Note that
$\gamma_{\mu\nu}=e^{-2\sigma(y)}\eta_{\mu\nu}$ is the background
spatial metric.

Define the five dimensional retarded Green's function, which satisfies
\begin{eqnarray}
\left[e^{2\sigma(y)}~\Box^{(4)} + \partial_y^2
-4k^{2}+4k(\delta(y)+\delta(y-a) )\right] G_R(x,x')~= \delta^{(5)}(x-x').
\label{greenfunct}
\end{eqnarray}

The formal solution of (\ref{rseomgr}) is then given by
\begin{equation}
 h_{\mu\nu}(x)=-2\kappa \int d^5 x' ~G_R(x,x')~\Sigma_{\mu\nu}(x')~
\delta(y-~a),
\label{formalsol}
\end{equation}
where integration is taken over the corresponding surface for which
$h_{\mu \nu}$ is to be determined. Since $h^{\mu}{}_{\mu}$ must vanish, 
we must impose $\Sigma^{\mu}{}_{\mu}=0$, 
which implies the ``equation of motion'' for $\xi^5$.

\begin{equation}
\Box^{(4)}~ \xi^5 = {\kappa \over 6} T \label{ywall}
\end{equation}
With this choice of $\xi^5$, it is easy to check 
that $h_{\mu\nu}$ given by eq.(\ref{formalsol}) satisfies
the harmonic condition  $h_{\mu}{}^{\nu}{}_{,\nu}=0$. The metric
perturbation on the brane can be decomposed into pieces induced by
standard model fields on the brane and the deformation of the brane
itself, as \cite{GT}
\begin{equation}
 h_{\mu\nu}= h^{(m)}_{\mu\nu}
 +2 k~\gamma_{\mu\nu}\xi^5,
\label{barhsimple}
\end{equation}
where, 
\begin{eqnarray}
&& h^{(m)}_{\mu\nu}=-2\kappa \int d^4 x' ~G_R(x,x') 
~\left(T_{\mu\nu}-{1\over 3}\gamma_{\mu\nu} T\right)(x'),
\label{onethird2}
\\
&& h^{(\xi)}=-4\int d^4 x'~ G_R(x,x')~\xi^5(x').
\label{onethird}
\end{eqnarray}
Here in (\ref{onethird2}), $h^{(m)}_{\mu\nu}$ corresponds
to the metric fluctuation induced by matter fields and $ h^{(\xi)}$ in
(\ref{onethird}) corresponds to that induced by brane deformation.

Now, let us look into the solution of eq. (\ref{greenfunct}). To
dimensionally reduce to four dimensions, the five dimensional graviton
field is expanded in Kaluza-Klein modes as
\begin{equation}
 \ h_{\mu \nu}( x, y) = \sum_{p=0}^{\infty} h_{\mu \nu}{}^{(p)}
(x)~ \psi^{p}(y) \label{decog}
\end{equation}
Defining
$z \equiv sgn(y) \left( e^{\sigma(y)} -1 \right)/k$,
$\psi_{p}(z) \equiv \psi_{p}(y) e^{\sigma(y)/2}$  ,  $h_{\mu \nu}(x,z)
\equiv h_{\mu \nu}(x,y) e^{\sigma(y)/2}$, the Kaluza Klein modes of
the graviton field can be expressed as solutions to an
effective one dimensional Schr\"odinger problem 
\begin{equation}
\left [\partial_z^2 - V(z) \right ]\psi_{p}(z)  = m^2 \psi_{p}(z),
\label{modeeq}
\end{equation}
with a potential
\begin{equation}
V(z)={15 k^2 \over 4 (k|z|+1)^2} -{3 k} \delta (z) +\frac{3 k}{1+ k
z_{r}} \delta (z- z_{c}).
\end{equation}

Our interest is in the zero mode which corresponds to the solution
$\psi_{0}(z) = N_0 (k |z| + 1)^{-3/2}$, where, $N_0$ is a normalization
constant. It is easily seen that that this mode satisfies boundary
conditions  required for the self
adjointness of the differential operator in eq.(\ref{modeeq})
\begin{eqnarray}
\partial_z {\psi}(0) &~=~& - {3 k \over 2 } \psi(0) ~\label{rsgrbc1}  \cr
\partial_z {\psi}(z_c) &~ =~& - {3 k \over 2 (k z_c +1)}
{\psi}(z_c)~.~\label{rsgrbc2}
\end{eqnarray}
where, $z_c \equiv (e^{\sigma(a)} - 1)/k$.
The modes satisfy the orthogonality condition
\begin{eqnarray}
\int_{-a}^{a} {e^{2\sigma(y)}}\psi_{p}(y) \psi_{q}(y) dy= \delta_{p q}
~\label{rsorthog}
\end{eqnarray}
which fixes the value of $\psi_{0}(y) = \sqrt
{k} e^{-2\sigma(y)}/(1- e^{-2\sigma(a)}) ^{1 \over 2}$. 

Notice that if we want to have a four dimensional Lagrangian for
$h_{\mu \nu}(x, y)$  with a canonical kinetic term and 
with masses of the $h_{\mu \nu}(x, y)$ field coming from the
Kaluza- Klein mode, we must have $dim[h]~=~3/2$, since
$h_{\mu \nu}(x, y)$ is a $5$  dimensional field. Also further since
$\psi$ is of dimension $1/2$, $h_{\mu \nu}(x)$ is of dimension $1$  as
expected. But, since $\eta_{\mu \nu}$ is dimensionless,
$h_{\mu \nu}(x, y)$ must be dimensionless. We choose to make the modes
of  $h_{\mu \nu}(x, y)$
dimensionless  by scaling $ h_{\mu \nu}(x)$ by $M_{P}$. Essentially,
we scale each  $h_{\mu \nu}(x, y)$ by  $M_{5}^{3 \over
2}~=~M_{P}~k^{1 \over2}$.

Thus, in terms of the redefined fields, we have 
\begin{equation}
\psi_{0}(y) = e^{-2\sigma(y)}/(1- e^{-2\sigma(a)})
\end{equation}
The Green's function is then
constructed from the complete set of eigenfunctions
\begin{equation}
G_R(x,x')=-\int {d^4 p\over (2\pi)^4}
e^{ip_{\mu}(x^{\mu}-x'{}^{\mu})}\Biggl[ {e^{2\sigma(y)} e^{2\sigma(y')}
\over 
{\bf p}^2-(\omega+i\epsilon)^2}{1\over (1- e^{-2\sigma(a)})} +\int_0^{\infty}
dm\, {u_m(y) u_m(y')\over m^2+{\bf p}^2-(\omega+i\epsilon)^2}\Biggr],
\end{equation}
where the first term corresponds to the zero mode and the rest
corresponds to the continuum of KK modes $ u_m(y)=\sqrt{m/2k}$
$\{J_1(m/k)Y_2(m/ka) -  Y_1(m/k) J_2(m/ak)\}$ $
/\sqrt{J_1(m/k)^2+Y_1(m/k)^2}$.

The Green's function for the zero mode contribution for
the negative and positive tension brane are given respectively as  
\begin{eqnarray}
&&G_R(x,x')~=~{ { e^{4\sigma(a)} \delta^{(4)}(x^{\mu}-x{}'^{\mu})} \over
{\Box^{(4)}(1- e^{-2\sigma(a)})}}\nonumber\\
&&G_R(x,x')~=~{{  \delta^{(4)}(x^{\mu}-x{}'^{\mu})} \over
{\Box^{(4)}(1- e^{-2\sigma(a)})}}
\end{eqnarray}

We find that in the zero mode
approximation the gravitational field on each  of the branes satisfies
\cite{GT}
\begin{equation}
\left({e^{2\sigma(y)}}~{}\Box^{(4)} h_{\mu\nu}\right)^{(\pm)} =
-\sum_{\alpha=\pm} 16\pi G^{(\alpha)} \left(T_{\mu\nu}-{1\over 3}
\gamma_{\mu\nu} T\right)^{(\alpha)}  \pm {16\pi G^{(\pm)}\over
3}{\sinh(ak)\over e^{\pm ak}}  \gamma_{\mu\nu} T^{(\pm)},
\label{big}
\end{equation}
where the plus and minus refer to quantities on the brane with positive
and negative tension respectively.  Here, we have introduced
\begin{equation}
G^{(\pm)}={G_5  e^{\pm ak}\over 2 \sinh(ak)}
\end{equation}
which plays the role of Newton's constant in a Brans-Dicke
parameterization \cite{GT}.

Let us now look at the KR - gravitational interaction as observed on
the visible brane. The equation (\ref{krbii}) becomes,
\begin{eqnarray}
\Box_{RS} V^M ~=~ M_5^{-3/2}~^*R^{MNPQS}~R_{NPQS} ~, \label{grrskr}
\end{eqnarray}

The axion coupling is
\begin{eqnarray}
\Box_{RS} \Phi_H ~=~M_5^{-3/2}~\left[R^{\mu}{}_{\nu \lambda \sigma
}~^*R_{\mu}{}^{\nu \lambda \sigma }~+ 2 \epsilon^{\mu \nu \lambda
\sigma}R_{\lambda \sigma}{}^{\delta 4}R_{\mu \nu \delta 4}\right] ~|_{y=a} ~
. \label{maxg}
\end{eqnarray}
For the zero modes of the fields, at the negative tension brane, we
obtain 
\begin{eqnarray}
\Box \Phi_H^{(0)}(x)~=~{\exp {\sigma(a)} \over {\sqrt{6}~M_P}}~ h_{\rho
 \eta}(x),_{\lambda}{}_{\sigma}[h_{\beta}{}^{\rho}(x),{}^{\lambda}{}_{\alpha}
 - h_{\beta}{}^{\lambda}(x),{}^{\rho}{}_{\alpha}]  e^{ \alpha \beta
   \sigma \eta} ~,\label{phieq}
\end{eqnarray}

For the vector field, the corresponding equation yields
\begin{eqnarray}
\Box V^{\mu}{}^{(0)}(x)~=~4\sqrt{2}{k \over M_P }~
 h_{\nu \beta}(x),_{\alpha}[h_{\lambda}{}^{\beta}(x),{}_{\sigma}{}^{\alpha}
 - h_{\lambda}{}^{\alpha}(x),{}^{\beta}{}_{\sigma}]  e^{ \mu \nu
\lambda \sigma} ~,\label{veceq}
\end{eqnarray}

Note that while the axion coupling to the graviton is enhanced by 
an exponential anti-warping factor, the vector field coupling shows no
such enhancement. Now, let us return to the equation of motion for the
graviton and consider the case where, $ka$ is large, so that
$1- e^{-2ka} \approx 1$. Then from (\ref{big}), on the negative tension
brane the equation reduces to\footnote{One could also start from
eq. (\ref{big}) but since the result will not change under such a
consideration, we prefer to get the result more simply.}
\begin{equation}
- \Box^{(4)} h_{\mu\nu}=~{8\pi e^{-2\sigma(a)} G_{5} \over 3}~T_{\mu \nu} 
\label{reduced}
\end{equation}

Let us now look at the energy-momentum tensor in
eq. (\ref{reduced}). Under a linear approximation, we get
\begin{eqnarray}
&&T_{\mu \nu}~=~ t_{\mu \nu} - {8\sqrt{6k}e^{5\sigma(a)}\over
M_5^{3 \over 2}}~\epsilon_{(\mu|}^{~~~\sigma \alpha
\beta}~\left[\Phi_{H}(x){}_{,
\lambda \sigma}~\left(h_{\beta |\nu)}(x),{}_{\alpha}^{~\lambda}~+
~h^{\lambda}_{~\beta}(x),{}_{\alpha |\nu)} \right) ~- ~\Phi_{H}(x)
{}_{,\alpha}~\Box
h_{\beta |\nu)}(x),{}_{\sigma} \right] ~\nonumber\\
&&~~~~~~~~~~~~~~~~~~~~~~~~~~~~~~~~~~~~~
~-{8k\sqrt{2k}e^{4\sigma(a)}\over
M_5^{3 \over 2}}~\epsilon_{(\mu|}^{~~~\sigma \alpha
\beta}~\left[V_{\alpha}(x),{}_{\sigma}~ h_{\rho \beta}(x),^{\rho}{}_{|\nu)}
+ V_{\alpha}(x),{}_{\sigma}{}^{\rho}~h_{\rho \beta}(x),{}_{|\nu)}\right]
\end{eqnarray}
In the above equation, we have used the mode solutions of the fields
on the brane. The energy momentum tensor $t_{\mu \nu}$
can be calculated from (\ref{emtensor})
by using (\ref{krsol}). Then, under the assumption of a
homogeneous axion background $\Phi^{(0)}_{H}(x)= \Phi^{(0)}_{H}(t)$
such that
$d\Phi_{H} ^{(0)} (x)/dt~=~f_{0}$ and homogeneous vector field,
restricting to the lowest order in Planck mass and $O(h)$ for
which a nontrivial effect is observed and hence ignoring
terms in the R.H.S. of eq. (\ref{phieq}) and eq. (\ref{veceq}), the 
equation of motion for the graviton turns out to be
(with the relevant terms in view of above
approximation)

\begin{eqnarray}
&&\Box^{(4)}~ h_{\mu\nu} = {8 \over M_{P}^{2}}(\partial_{\sigma}
\Phi_{H}(x)~  \partial_{\eta} \Phi_{H}(x) ~\eta^{\sigma
\eta})~(\eta_{\mu \nu} + h_{\mu \nu} (x)) +{4 \over 3M_{P}^{2}}
(\partial_{\sigma} V_{\eta}(x)~ \partial_{\alpha}V_{\beta}(x)~
\eta^{\sigma \alpha}~ \eta^{\eta \beta})~(\eta_{\mu \nu} + h_{\mu
\nu}(x))\nonumber\\ &&~~~~~~~~~~~~~+{8\sqrt{6}e^{3\sigma(a)}\over 3
M_P^{3}}~\epsilon_{(\mu|}^{~~~\sigma \alpha
\beta}~\left[\Phi_{H}(x){}_{, \lambda \sigma}~\left(h_{\beta
|\nu)}(x),{}_{\alpha}^{~\lambda}~+ ~h^{\lambda}_{~\beta}(x),{}_{\alpha
|\nu)} \right) ~- ~\Phi_{H}(x) {}_{,\alpha}~\Box h_{\beta
|\nu)}(x),{}_{\sigma} \right] ~\nonumber\\
&&~~~~~~~~~~~~~~~~~~~~~~~~~~~~~~~~~~~~~~~~~~~~~~~~~~~~
+{8\sqrt{2}k\over 3M_P^{3
}}~\epsilon_{(\mu|}^{~~~\sigma \alpha
\beta}~\left[V_{\alpha}(x),{}_{\sigma}~ h_{\rho
\beta}(x),^{\rho}{}_{|\nu)} +
V_{\alpha}(x),{}_{\sigma}{}^{\rho}~h_{\rho \beta}(x),{}_{|\nu)}\right]
\end{eqnarray}

Proceeding as in the section III, 
the circularly polarized components lead to the equation\footnote {We are
interested only in the axion coupling. Also the KR vector coupling
does not lead to the rotation of polarization plane of the
gravitational wave. We will ignore the KR vector-graviton coupling here.}
\begin{eqnarray}
\left[{d^2 \over dt^2}~+~p^2 ~+~{\cal F}_{\pm} \right]
\varepsilon_{\pm}~=~-~{\cal F}_{\pm}~, \label{dife}
\end{eqnarray}
where,
\begin{eqnarray}
{\cal F}_{\pm}~\equiv~{8\pi f_0^2  \over M_P^2~(1 \pm  {64
    \sqrt 6 \pi p f_0 e^{3\sigma(a)} /M_P^3}) } ~ , \label{eff}
\end{eqnarray}
and
\begin{eqnarray}
h_{ij}~=~\varepsilon_{ij}(t)~\exp ~-ipz ~. \label{rsgw}
\end{eqnarray}

Even though we are interested in large $p$, we would still remain
within the Planckian regime $p < M_P$ so that the quantity $64\sqrt6\pi p
f_0e^{3\sigma(a)}~/M_P^3 << 1$ and can serve as an expansion parameter, 
leading to
\begin{eqnarray}
\left[{d^2 \over dt^2} ~+~ p^2 ~+~ 8\pi f_0^2 / M_P^2  ~\mp~512
\sqrt 6\pi^2 p f_0^3 e^{3\sigma(a)} /M_P^5 \right] \varepsilon_{\pm}~
\simeq~-~8\pi f_0^2~~(1 \mp 64\sqrt6
\pi p f_0 e^{3\sigma(a)}/M_P^3)/M_P^2~ . \label{diffe}
\end{eqnarray}
We can now read off the dispersion relation
\begin{eqnarray}
\omega_{\pm}^2 ~ = ~p^2 ~+~ 8\pi f_0^2 / M_P^2  ~\mp~512
 \sqrt 6\pi^2 p f_0^3 e^{3\sigma(a)} /M_P^5 \label{disp}
\end{eqnarray}
Therefrom, we calculate the group velocity
\begin{eqnarray}
v_{g \pm}~ \equiv ~ {d\omega_{\pm} \over dp} ~=~ 1 ~+~ O(p^{-2})~,
\label{gvel}
\end{eqnarray}
and the phase velocity
\begin{eqnarray}
v_{p \pm}~ \equiv ~ {\omega_{\pm} \over p} ~=~ 1\mp~256
 \sqrt 6\pi^2 p f_0^3 e^{3\sigma(a)} /M_P^5 .\label{rspvel}
\end{eqnarray}
Thus, for large $p$, the violation of Lorentz invariance can be
ignored. As in the electromagnetic case, the rotation of the
polarization plane for gravitational waves is RS- spacetime is given by
\begin{eqnarray}
\Delta \Psi_{grav}~\simeq~{512
\sqrt 6~\pi^2~ f_0^3 e^{3\sigma(a)} \over M_P^5}  ~~\Delta t ~=
{512 \sqrt 6~\pi^2~ \tilde f_0^3 \over M_P^5} ~~\Delta t . 
\label{rsgrota}
\end{eqnarray}
So, the rotation of the plane of polarization for the 
gravitational wave is enhanced by the anti-warping factor of $\exp 3
\sigma(a)$. Even with very conservative limits for $f_{0}$, the effect
becomes far larger than what anyone ever expects to observe. 


\section{Discussion}

The rotation of the plane of polarization of gravitational waves upon
interaction with a homogeneous Kalb-Ramond field background, is a new
aspect of Kalb-Ramond physics which has hardly been explored. The
exponential enhancement of the effect in an RS1 scenario is quite
analogous to the corresponding effect of KR fields on electromagnetic
waves from cosmologically distant sources. However, the huge
disparity between our theoretical range of values for the
angle of rotation of the plane of polarization, both for
electromagnetic and gravitational waves, and observational constraints
already available, points to serious flaws within the theoretical
framework used. The issue is then, precisely {\it where} are these
flaws ? The basic assumptions made in the work have already been
mentioned in an earlier part of the paper, but for completeness we
recapitulate them
\begin{enumerate}
\item Ten dimensional heterotic string theory admits the warped five
dimensional (RS1) spacetime as a consistent compactification. Note
that `RS1' for us means that the visible brane possesses negative
tension, as is crucial to produce a naturally light Higgs scalar.
\item The Kalb-Ramond field couplings to Maxwell fields and linear
metric fluctuations in this warped spacetime is the
straightforward dimensional reduction of the couplings in ten
dimensions in quantum heterotic string theory. 
\item The compactification preserves the gauge invariance of the five
dimensional Hodge-dual Kalb-Ramond vector field of which the axion is
the fourth component. 
\end{enumerate}
If any of these assumptions are incorrect, our results no longer
stand. On the other hand, if none is actually incorrect, our result
points to a fundamental tension between heterotic string theory and
the warping of spacetime introduced by \cite{rs} in their attempt to
resolve naturalness and gauge hierarchy issues in the standard
strong-electroweak theory. Put somewhat differently, reconciling the
results we have obtained with observational constraints requires fine
tuning of the constant $f_0$ to the same extent as one is forced to do
within the standard electroweak theory to obtain a naturally light Higgs
boson. Such a high degree of fine tuning is an extremely unsavoury
feature of a scheme meant to do away with the fine tuning problem in
the standard model. 

One may mention that the RS1 braneworld scenario, with the visible
brane having negative tension, has already been shown to lead to
inconsistencies with observational constraints \cite{GT}. The spin two
fluctuations on the visible brane lead to a Brans-Dicke type of
gravity with a Brans-Dicke parameter lying far beyond the range
restricted by observational constraints based on solar system tests.
Some have tried to argue away such inconsistencies by conjecturing
that the
effective Brans-Dicke scalar will, on embedding in some sort of a
string theory, turn out to be a moduli field which may acquire a
potential, and therefore no longer produce the offending effect. While
this is a distinct possibility, the crucial task of showing that the
vacuum which stabilizes the Brans-Dicke moduli field also has the
standard model fields with the correct spectrum, remains.

Why cannot a similar moduli stabilization work for the KR axion ?
Consistent with our assumptions, the KR axion is the fourth component
of a five-vector gauge field whose masslessness is protected by a dual
version of Kalb-Ramond gauge invariance. Stabilizing a one-off scalar
moduli may be easier than one which is a component of a massless gauge
field, since one now needs a Higgs mechanism to be simulated by
appropriate fluxes on the brane. Additionally, the problem of
demonstrating that the same Higgs vacuum yields the standard model
spectrum remains. Nothing so far is known about this latter
vacuum. Until such time as these problems are resolved, the fundamental
inconsistencies discerned above cannot be argued away.

\begin{acknowledgements} We thank P. Jain for interesting discussions. 
One of us (PM) thanks J. Maharana, S. Panda, S. SenGupta and S. Trivedi
for illuminating discussions, and especially R. Maartens for
discussions on ref.\cite{GT} and the solar system constraints on the
Brans-Dicke parameter. 
\end{acknowledgements}


\end{document}